 \documentclass[authoryear,preprint,review,12pt]{elsarticle}

\usepackage{amssymb}
\begin{document}

\begin{frontmatter}



\title{Fermi Statistics of Partially Populated States}


\author{R. A. Treumann}

\address{%International Space Science Institute, Bern, Switzerland \\
Department of Geophysics, Munich University, Munich, Germany}

\begin{abstract}
Fermi statistics is formally extended to the case when energy levels
are allowed to be partially occupied, which the Pauli principle does not categorically exclude.
The \emph{partial} Fermi distribution obtained depends on the partial occupation of states but otherwise
has similar properties as the (integer) Fermi distribution. In the zero temperature limit both are
identical.
\end{abstract}


\end{frontmatter}


%\section{}
%\label{}

There are three kinds of particles, Bosons, Fermions, and Anyons the latter obeying fractional statistics which is a mixture between bosonic and fermionic statistics. Bosons are allowed to occupy any energy states to arbitrary numbers. For Fermions, similar behavior is inhibited by the Pauli principle which apparently allows for only two occupations, empty states or (when neglecting the particle spin) one particle per state. Fermi's ingenious truncation of the infinite sum in the partition function $Z_i=\sum_i\{\exp[(\mu-\epsilon_i)/k_BT]\}^{n_i}$ by boldly assuming only binary occupations $n_i=[0,1]$ of energy states, $\epsilon_i$ with $i\in\textsf{N}$, had profound consequences for the statistical mechanics of solids at low temperatures. It immediately led him to the proposal  of his celebrated (Fermi) distribution $\langle n_i\rangle_F=\{1+\exp[-(\mu-\epsilon_i)/k_BT]\}^{-1}$. His assumption was justified by the Pauli principle and for spin-$\frac{1}{2}$ particles was ultimately given its quantum mechanical interpretation based on the complete asymmetry of Fermionic wave functions. Discovery of the quantum Hall effect, in particular the fractional effect, had temporarily shaken Fermi statistics, leading to suggestion of anyon statistics, until Laughlin's  \citep{laughlin1983} justifying proposal of his wave function which includes interaction with bosonic fields.  

What, on the other hand, happens, when sufficiently many states are available for Fermions (e.g. electrons) and the electrons would be allowed to fill such states only partially, i.e. in fractions? As for an example one may think of gyrating electrons which bounce in a magnetic mirror geometry at frequency $\omega_b\ll\omega_{ce}$. In this case Landau levels $\epsilon_L= \omega_{ce}\hbar (L +\frac{1}{2}), ~L\in \textsf{N}$, split into a number of bounce levels $\epsilon_b= \hbar\omega_b(b+\frac{1}{2}), ~ b\in\textsf{N}, ~b/L<\omega_{ce}/\omega_b$. The total electron energy $\epsilon_{b,L}=\epsilon_L+\epsilon_b$ in Landau level $L$ is then shared by the two kinds of oscillatory states of the electron. Under these circumstances, all the electron energy is in the Landau levels at the mirror points, while in the minima of the magnetic field a substantial part of energy is transferred to the bounce levels. Theoretically, bounce levels might then become only partially filled under these circumstances, even though only fermionic states are involved, and partial population of states may not necessarily mean that the Pauli principle is violated, when not involving bosonic interactions. The
occupation may still be a fraction below one which the Pauli principle not explicitly excludes. Though I am not aware of any observations of this case, in the following I rewrite the formalism for the partial population case.

This is easily done, starting as usually \citep[cf., e.g.,][]{huang1987,landau1994} from the logarithm $\Omega_i[n_i]$  of the $\Gamma$-phase space volume corresponding to the population numbers $[n_i]$ of the particles in an ideal gas:
\begin{equation}
\Omega_i = -k_BT\log \sum_{n_i} \left(\exp \frac{\mu -\epsilon_i}{k_BT}\right)^{n_i},
\end{equation}
with $\mu$ chemical potential, $k_BT$ temperature, both in energy units, $\epsilon_i=p_i^2/2m$ particle energy, $\mathbf{p}$ particle momentum, $m$ mass, and the sum over Gibbs distributions in states $n_i$ the canonical partition function $Z_i$. 

Assume that the states can become partially occupied by Fermions alone. Occupation numbers $n_i>1$ are excluded by the Pauli principle. Hence, partial population implies that, given the interval $[0,\ell]$ with fixed natural number $\ell\in\textsf{N}$, and $j$ any integer such that $j\in[0,\ell]$, the thermodynamic potential $\Omega_i$ can be written
\begin{equation}
\Omega_i=-k_BT\log\sum_{j=0}^\ell\left(\exp\frac{\mu-\epsilon_i}{\ell k_BT}\right)^j, \qquad j\in[0,\ell]
\end{equation}
where $j=0,\ell$ just reproduces the two Fermi occupations. In the intermediate interval the populations follow the simple partial chain $\{j/\ell\}$, with fixed $\ell\geq j$. Summation of the sum becomes simple matter since it represents a truncated geometric progression with ratio $\exp[(\mu-\epsilon_i)/\ell k_BT]$ yielding
\begin{equation}
\Omega_i=-k_BT\log\left\{\frac{\exp[x_i(\ell+1)]-1}{\exp(x_i) - 1}\right\}, \qquad x_i\equiv \frac{\mu-\epsilon_i}{\ell k_BT}.
\end{equation}
One trivially shows that this becomes the thermodynamic Fermi potential $\Omega_{iF}=-k_BT\log\left\{1+\exp\left[(\mu-\epsilon_i)/k_BT\,\right]\right\}$ for $\ell=1$. From here, taking the derivative $-\partial\Omega_i/\partial\mu$, the average \emph{partial} distribution $\langle n_i\rangle_\ell$ in the $i$th quantum state follows as 
\begin{equation}
\langle n_i\rangle_\ell =\frac{1}{\ell}\left[\frac{(\ell+1)\mathrm{e}^{x_i(\ell+1)}}{\mathrm{e}^{x_i(\ell+1)}-1}-\frac{\mathrm{e}^{x_i}}{\mathrm{e}^{x_i}-1}\right],\quad \ell\geq 1,\label{eq-fracferm}
\end{equation}
It is again easily shown that the ordinary Fermi distribution $\langle n_i\rangle_F$ is reproduced for $\ell=1$. 

The  partial Fermi distribution Eq. (\ref{eq-fracferm}) is a bit more complicated than the (integer) Fermi distribution. 
Apparently, it looks more like a Boson distribution. However, this is an illusion which becomes clear when checking its low and high temperature forms which, as they should, agree with those for the Fermi distribution. For $T\to 0$ one obtains $\mu=\epsilon=\epsilon_F$, $\langle n_i\rangle\to 1$. For $T\gg\mu\sim\epsilon_F$ one recovers the Boltzmann distribution.

Thus the fermionic property of the distribution is maintained. Obviously it results from the subtraction of two bosonic distributions in Eq. (\ref{eq-fracferm}). The anti-symmetric property of the partial many-particle fermionic system being caused by subtraction. 

From the partial distribution $\langle n_i\rangle_\ell$ as function of energy $\epsilon_i(\mathbf{p})$ of momentum $\mathbf{p}$ all thermodynamic quantities like the equation of state can be derived formally defining the appropriate moments. For an ideal Fermi gas one then has for the pressure $P$ and density $N$, respectively
\begin{equation}
\frac{P}{k_BT}=\frac{1}{\lambda_T^3}f_\ell(z), \qquad N=\frac{1}{\lambda_T^3}z\partial_zf_\ell(z)
\end{equation}
with $\log z = \mu/k_BT$, when introducing an appropriate new function
 %One may also write down the expressions for the equation of state and density of an ideal partial Fermi gas by replacing $\langle n_i\rangle$ with its momentum dependent form $\langle n_\mathbf{p}\rangle$, i.e. $x_i\to x_\mathrm{p}=(\mu-\epsilon_\mathrm{p})/\ell k_BT$,  and integrating over all momenta $\mathbf{p}$. The gas pressure of the grand canonical ensemble is defined as $PV=-\sum_i\Omega_i$. Transforming the sum into a momentum integral, $\Omega_i\to\Omega_\mathbf{p}$ depends on the momentum. For the partial Fermi case, this yields the ideal gas equation of state
%\begin{eqnarray}\label{eq-state}
%\frac{PV}{k_BT}&=&\frac{1}{2\pi^2\hbar^3}\int_0^\infty p^2\mathrm{d}p\,\log\left(\frac{\mathrm{e}^{x_\mathbf{p}\ell}-1}{\mathrm{e}^{x_\mathbf{p}}-1}\right), \\
%\frac{N}{V}&=&\frac{1}{2\pi^2\hbar^3}\int_0^\infty d^2\mathrm{d}p \langle n_\mathbf{p}\rangle,
%\end{eqnarray}
%where $\langle n_\mathbf{p}\rangle$ is given in Eq. (\ref{eq-fracferm}). With fugacity $z= \exp(\mu/k_BT)$, and since $\langle n_\mathbf{p}\rangle$ is the derivative of $\Omega_\mathbf{p}$, and with the square of the  thermal de Broglie wavelength $\Lambda^2={2\pi\hbar^2/mk_BT}$, the equation of state and density can then be given their usual forms
%\begin{equation}
%\frac{PV}{k_BT}=\frac{f_\ell(z)}{\Lambda^3}, \qquad \frac{N}{V}=\frac{z}{\Lambda^{3}}\frac{\partial f_\ell(z)}{\partial z},
%\end{equation}
\begin{equation}\label{eq-fell}
f_\ell(z)=\frac{4}{\sqrt{\pi}}\int\limits_0^\infty y^2\mathrm{d}y\log\left(\frac{1-z^{1+\frac{1}{\ell}}\mathrm{e}^{-\left(1+\frac{1}{\ell}\right)y^2}}{1-z^{1/\ell}\mathrm{e}^{-y^2/\ell}}\right)\approx \ell^\frac{3}{2}\sum\limits_{k=1}^\infty \frac{(-1)^{k+1}z^{k/\ell}}{k^{5/2}}
\end{equation}
which replaces the usual function $f_{5/2}(z)$. Here, $\lambda_T=\sqrt{2\pi\hbar^2/mk_BT}$ is the thermal wavelength. As usual, ${\cal N}=z\partial_zf_\ell(z)$ is the particle number per thermal wavelength volume.

One may even formally extend the partial case to the extreme case of a continuity of partial states. Then the sum in the expression for the thermodynamic potential $\Omega_i$ turns into an integral yielding that
%\begin{equation}
%\Omega_i=-k_BT\log \int_0^1 ds\,\exp\left(\frac{\mu-\epsilon_i}{k_BT}s\right),
%\end{equation}
%which immediately yields that
\begin{equation}
\Omega_i = -k_BT \log \left\{ \frac{k_BT}{\mu-\epsilon_i} \left[ \exp \left( \frac{\mu-\epsilon_i}{k_BT} \right) -1\right] \right\}.
\end{equation}
and for the average distribution
\begin{equation}
\langle n_i\rangle = \frac{\exp\left[(\mu-\epsilon_i)/k_BT\right]}{\exp\left[(\mu-\epsilon_i)/k_BT\right]-1}-\frac{k_BT}{\mu-\epsilon_i}.
\end{equation}
It is straightforward to show by expanding that, in the limit $k_BT\to 0$, it becomes
\begin{equation}
\langle n_i\rangle \simeq 1-k_BT/(\mu-\epsilon_i), \qquad \epsilon_i<\mu,
\end{equation}
%Again, as a consequence, for $\mu<\epsilon_i$ the distribution vanishes with vanishing temperature. 

%The equation of state in this case assumes the following form
%\begin{equation}
%\frac{PV}{k_BT}=\frac{4}{\sqrt{\pi}\Lambda^3}\int\limits_0^\infty y^2\mathrm{d}y\log\left(\frac{z\mathrm{e}^{-y^2}-1}{\log (z) -y^2}\right)
%\end{equation}

In summary, from the purely statistical mechanical point of view there is no obvious contradiction between the Pauli principle for Fermions and a partial occupation of states. Statistical mechanics of Fermions, respecting the Pauli principle, does  not categorically exclude the existence of partially occupied states as long as the partial occupation number remains to be smaller than unity. 

The partial Fermi distribution Eq. (4) is a variant of the (integer occupation) Fermi distribution which it reproduces for $\ell=1$. The low and high temperature limits are the same, as is the definition of Fermi energy. Fermions occupying partial states at $T=0$ remain to be degenerate, even for the case of continuous occupation. 

The present investigation so far lacks any application as well as its quantum mechanical justification. The assumed partial occupation is solely fermionic with no Bosons involved. This differs, for instance, from the Quantum Hall effect. The restriction to Fermions naturally implies a somewhat different physics, if partial occupations should in some way be quantum mechanically realized. It, however, shows that statistical mechanics alone does not a priori inhibit purely fermionic partial occupations.

\subsection*{Acknowledgement}
{Hospitality of the International Space Science Institute, Bern, during a one day visit is acknowledged.}




\begin{thebibliography}{99}

%\bibitem[1]{wilczek1982}
%F. Wilczek, Phys. Rev. Lett.\ {\bf 49}, 957 (1982)

%\bibitem[2]{haldane1983}
%F. D. M. Haldane, Phys. Rev. Lett.\ {\bf 51}, 605 (1983)

%\bibitem[3]{haldane1991}
%F. D. M. Haldane, Phys. Rev. Lett.\ {\bf 67}, 937 (1991)

%\bibitem[4]{wilczek1984}
%D. Arovas, J. R. Schrieffer, and F. Wilczek, Phys. Rev. Lett.\ {\bf 53}, 722 (1984)

%\bibitem[5]{halperin1984}
%B. I. Halperin, Phys. Rev. Lett.\ {\bf 52}, 1583 (1984)

%\bibitem[6]{wilczek1985}
%D. P. Arovas, R. Schrieffer, F. Wilczek, and A. Zee, Nuclear Phys. B {\bf 251}, 117 (1985)

%\bibitem[7]{wu1994}
%Y.-S. Wu, Phys. Rev. Lett.\ {\bf 73}, 922 (1994)

\bibitem[Huang(1987)]{huang1987}
Huang K., {\it Statistical Mechanics}, 2nd Edition (John Wiley \& Sons, New York, 1987).

\bibitem[Landau \& Lifschitz(1994)]{landau1994}
Landau L. D. \& Lifshitz E. M., {\it Statistical Physics}, 3rd Edition, Part 1 (Pergamon Press, Elsevier Science, Oxford, 1994).

\bibitem[Laughlin(1983)]{laughlin1983}
Laughlin R. B., Phys. Rev. Lett.\ {\bf 50}, 1392-1395 (1983).


\end{thebibliography}
\end{document}